\documentclass[twocolumn]{article}
\usepackage{graphicx}
\usepackage{amsmath,amssymb}

\title{Near ML detection using Dijkstra's algorithm with bounded list size over MIMO channels}
\author{Atsushi OKAWADO, Ryutaroh MATSUMOTO\\ and Tomohiko UYEMATSU\\
Dept.\ of Communications and Integrated Systems\\
Tokyo Institute of Technology, 152-8550 Japan}
\date{February 13, 2008}

\begin{document}
\maketitle
\begin{abstract}
We propose Dijkstra's algorithm with bounded list size 
after QR decomposition for decreasing the computational complexity 
of near maximum-likelihood (ML) detection of 
signals over multiple-input-multiple-output (MIMO) channels. After that, we compare 
the performances of proposed algorithm, QR decomposition M-algorithm (QRD-MLD), 
and its improvement. 
When the list size is set to achieve the almost same symbol error rate (SER)  as the 
QRD-MLD, the proposed algorithm has smaller average computational complexity.
\end{abstract}

\section{Introduction}
The channel capacity of multiple-input-multiple-output (MIMO) channels 
linearly increases with the number of antennas \cite{MI,MO}. 
Maximum-likelihood (ML) detection provides the  minimum error rate. 
However, the computational complexity of the simple ML detection 
algorithm grows exponentially with the number of transmit antennas. 
Thus, we need an efficient algorithm that achieves similar error 
rate to the ML detection. 
The QR decomposition M-algorithm (QRD-MLD) \cite{M1,M2} 
and sphere decoding (SD) \cite{SD} are possibly the most promising algorithms. 
In \cite{fukatani}, to reduce the computational complexity, 
Dijkstra's algorithm is applied to SD which 
achieves same error rate as ML detection. 
Both the QRD-MLD and Dijkstra's algorithm are tree search based algorithms. 
Dijkstra's algorithm uses the list of unlimited size to keep detection candidates. 
However, the computational complexities of the QRD-MLD and 
Dijkstra's algorithm are still high. 
To reduce the computational complexity, we propose Dijkstra's algorithm with bounded 
list size. When proposed algorithm's list size is set 
to achieve the almost same symbol error rate (SER)  as the QRD-MLD, 
the computational complexity of proposed algorithm 
is lower than the QRD-MLD. 

This paper is organized as follows. In Section 2, we introduce the 
system model of MIMO channels. In Section 3, we review the QRD-MLD 
and its improvement, then propose Dijkstra's algorithm with bounded list size. 
In Section 4, we show the comparison between the computational complexity 
of the QRD-MLDs and proposed algorithm by computer simulations. 
Finally, we give the conclusion in section 5.

\section{System model}

We consider the uncoded system with $t$ transmit antennas and 
$r$ receive antennas, and we assume $r \geq t$. We assume that the noise at each receive antenna
 is the additive white Gaussian noise (AWGN). Let $\texttt{x}$ be a $t \times 1$ 
vector consisting of complex envelopes of transmitted signals with the 
signal constellation $\texttt{S}$, $\texttt{H}$ an $r \times t$ fading matrix
whose $(k,j)$ entry is a complex fading coefficient between $j$-th transmit
 antenna and $k$-th receive antenna, $\texttt{z}$ an $r \times 1$ complex 
vector whose component is noise at each receive antenna, and $\texttt{y}$ 
an $r \times 1$ complex vector whose component is the received signal 
component at each receive antenna. The model of this channel is written as 
\begin{equation}
\texttt{y} = \texttt{Hx} + \texttt{z}  .
\end{equation}
We assume that the receiver knows the channel state information \texttt{H} perfectly.

In this case, the ML detection of the transmitted signal over the channel 
(1) can be formulated as finding 
\begin{equation}
\hat{\texttt{x}}_{ml} = \arg \min_{\texttt{x} \in \texttt{S}^t} || \texttt{y} - \texttt{Hx} ||^2 .
\end{equation}

\section{Near ML detection algorithm}
In this section, we propose the new 
near ML detection algorithm. 
First, to calculate (2) efficiently, we explain how to find the ML signal 
 by tree search algorithm 
in Section 3.1. Then, we review near ML detection algorithms 
called QRD-MLD \cite{M1,M2} and its improvement \cite{M3} in Section 3.2. 
Finally, we propose 
Dijkstra's algorithm with bounded list size in Section 3.3.

\subsection{QR decomposition}
To calculate (2) efficiently, 
we compute a QR decomposition of $\texttt{H}$ and obtain an upper triangular 
matrix $\texttt{R}$ and a unitary matrix $\texttt{Q}$ with $\texttt{H} = \texttt{QR}$.
Since $\texttt{Q}$ is unitary, 
\begin{equation}
|| \texttt{y} - \texttt{Hx} ||^2 = || \texttt{Q}^{\ast} \texttt{y} - \texttt{Q}^{\ast} 
\texttt{Hx} ||^2 = || \texttt{Q}^{\ast} \texttt{y} - \texttt{Rx} ||^2 .
\end{equation}
Let $\xi = \texttt{Q}^{\ast} \texttt{y} = (\xi_1, \cdots, \xi_r)^T$. The ML detection problem (2) can be reformulated as finding 
\begin{eqnarray*}
\hat{\texttt{x}}_{ml} & = & \arg \min_{\texttt{x} \in \texttt{S}^t} || \xi - \texttt{Rx} ||^2 \\
& = & \arg \min_{\texttt{x} \in \texttt{S}^t} \left \{ \sum_{j = 1}^{t} | \xi_j - \sum_{i = j}^{t} \texttt{R}_{j,i}x_i |^2 + \sum_{k = t}^{r} | \xi_k |^2 \right \} 
\end{eqnarray*}
\begin{equation}
= \qquad \arg \min_{\texttt{x} \in \texttt{S}^t} \left \{ \sum_{j = 1}^{t} | \xi_j - \sum_{i = j}^{t} \texttt{R}_{j,i}x_i |^2 \right \} .
\end{equation}
The second equality above follows as the second term in the second equation 
is irrelevant to $\texttt{x}$.

To calculate (4) efficiently, we consider a weighted directed graph as follows. 
The decisions on $x_i$ construct a tree where nodes at 
$k$-th depth are correspond to the candidate of $x_{t-k+1}$ \cite{tree}, 
and the root node is placed at depth 0. 
Then, the metric value, which is the weight of branch, between a node $\hat{x}_i$ that 
has $\hat{x}_t, \cdots, \hat{x}_{i+1} (\hat{x}_k \in \texttt{S}, i+1 \leq k \leq t)$ 
as ancestor nodes from the root node to its parent node is defined by 
\[ m_i = | \xi_i - R_{i,i} \hat{x}_i - \sum_{j = i+1}^{t} R_{i,j} \hat{x}_j |^2 . \] 
The distance of each node from the root node, which is called the 
accumulated metric value in this paper, is equal to the sum of the 
metric values of branches from the root node to the node itself. 
The accumulated metric value from the root node to the bottom node 
whose depth is $t$ is 
\begin{equation}
\sum_{i =1}^{t} m_i = \sum_{j = 1}^{t} | \xi_j - \sum_{i = j}^{t} \texttt{R}_{j,i} \hat{x}_i |^2 .
\end{equation}
Because $\hat{\texttt{x}}$ that makes (5) minimum is equal to $\hat{\texttt{x}}_{ml}$ of (4), 
the shortest path from the root node to the bottom node 
corresponds to  the ML signal \cite{tree}. 

\subsection{QRD-MLD}
The QRD-MLD \cite{M1,M2}, which is a breadth-first tree search 
based algorithm, finds a near ML signal. 
The QRD-MLD keeps only $M$ nodes at each depth with the 
smallest accumulated metric values \cite{M}, instead of testing all the candidate in  $\texttt{S}^t$ 
according to (4). At each depth, only $M$ nodes make 
their child nodes. We call a node that makes its child node detection node in this paper. 

An improvement to QRD-MLD proposed in \cite{M3} reduces the number of detection 
nodes from the original QRD-MLD. 
This improved QRD-MLD has threshold value at each depth. The depth $i$ 's 
threshold value $\Delta_i$ is defined by 
\begin{equation}
\Delta_i = E_{i,min} + X \phi^2 , 
\end{equation}
where $E_{i,min}$ is the smallest accumulated metric value of the node at 
$i$-th depth in the nodes whose parent node is a detection node. 
$X$ is a fixed constant number, and $\phi^2$ is the noise variance.
At each depth, select
the nodes that have smaller accumulated metric value than threshold value $\Delta_i$. 
If the number of selected nodes is more than $M$, only $M$ nodes with 
smallest accumulated metric values are selected.

Note that both algorithms do not always find the ML signal. 
For small to medium $M$ values, the complexity is substantially lower 
than the simple ML detection algorithm. However, the final result is no longer guaranteed to be the 
ML signal. 

\subsection{Proposed algorithm: Dijkstra's algorithm with bounded list size}
Dijkstra's algorithm is an efficient algorithm to find the shortest path 
from a point to a destination in a weighted graph \cite{D1}. 
Dijkstra's algorithm uses the list of unlimited size to keep candidate nodes. 
If we use Dijkstra's algorithm to find the shortest path from the root 
to one of nodes at the bottom depth, we can get the node with minimum 
$|| \texttt{y} - \texttt{H}\hat{\texttt{x}} ||^2$ among all nodes 
at the bottom depth and it corresponds to the ML estimate \cite{fukatani}.
However, this algorithm still has high computational complexity. To reduce the 
computational complxity, we propose a modified version of Dijkstra's algorithm 
whose list  keeps only $L$ nodes with the smallest 
accumulated metric values in the list. 

We show Dijkstra's algorithm with bounded list size.
\begin{enumerate}
\item Create an empty list for nodes. 
\item Insert all nodes at the first level into the list. 
\item Select the node A having smallest accumulated metric value in the list and remove it from the list. 
If the depth of A is $t$, then output the node A and its ancestor nodes 
as the ML signal and finish this algorithm. 
\item Insert all A's child nodes into the list. 
\item Arrange the nodes in the list according to the accumulated 
metric value by the quick sort. 
If the list has more than $L$ nodes, 
select the $L$ nodes with the smallest accumulated metric values in the list, and discard 
other nodes from the list.
\item Go back to Step 3.
\end{enumerate}

The node whose child nodes are inserted into the list is called 
detection node in this paper. 
Because 
the discarded nodes, which are decided at Step 5, and their descendant nodes 
are not examined, the proposed algorithm dose not examine all the candidate in 
$\texttt{S}^t$ according to (4). Thus, the proposed algorithm dose not always find the ML signal.

When we use LDPC codes \cite{LDPC} 
or turbo codes \cite{turbo} after detection, we 
have to compute $N$ most likely signals \cite{9}. 
Such signals can be computed by this algorithm's modification  
that is finished after output $N$ signals with the smallest 
accumulated metric value.

\section{Computer simulation}

\begin{figure}[t!]
\includegraphics*[width=\linewidth]{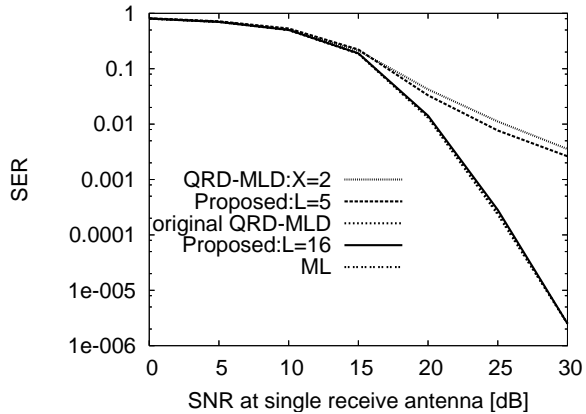}
\caption{($4 \times 4$) symbol error rate}
\end{figure}
\begin{figure}[t!]
\includegraphics*[width=\linewidth]{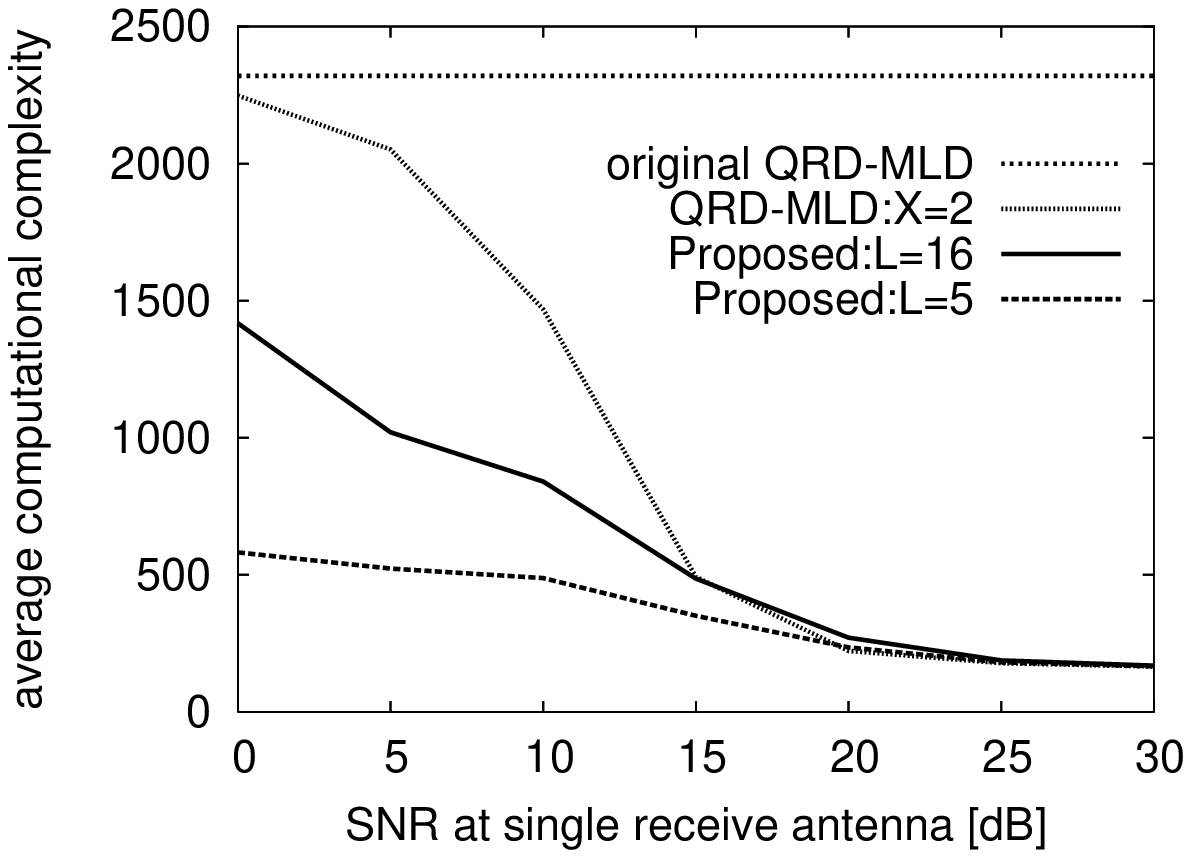}
\caption{($4 \times 4$) average computational complexity}
\includegraphics*[width=\linewidth]{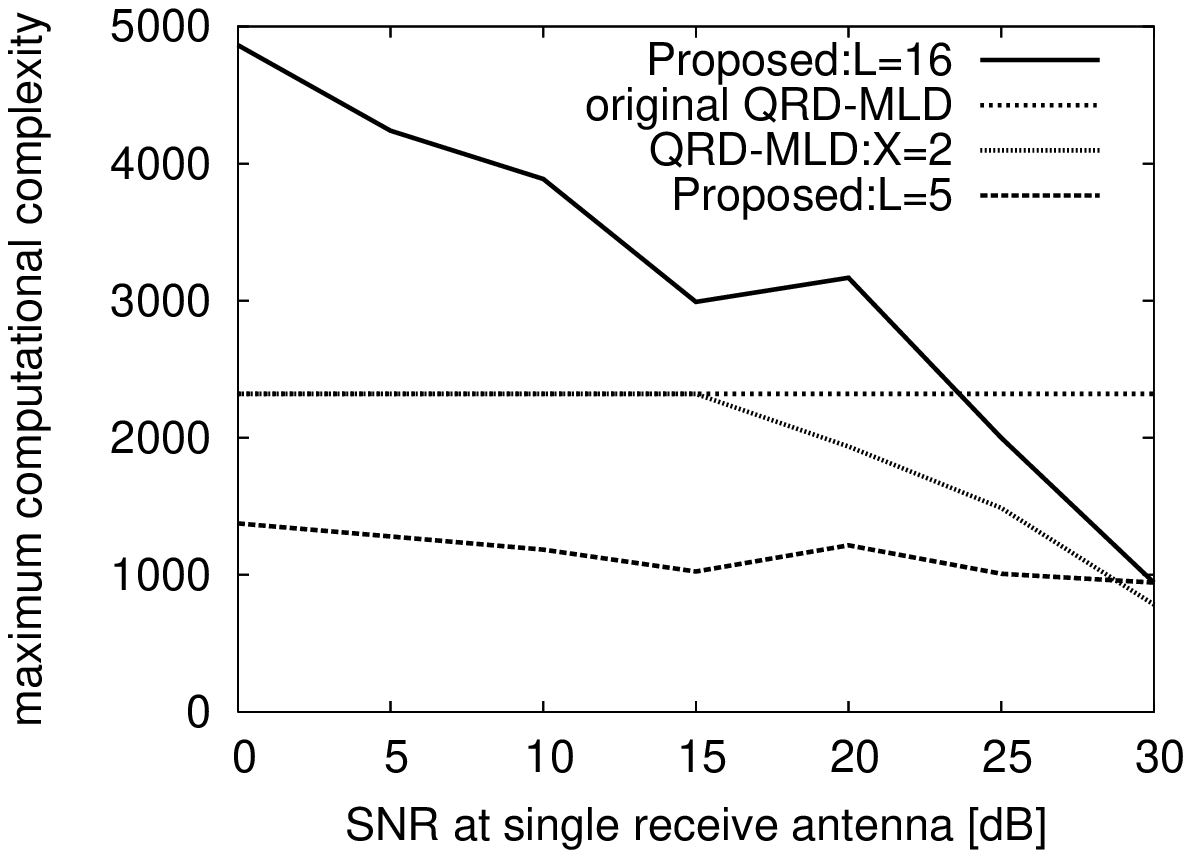}
\caption{($4 \times 4$) maximum computational complexity}
\end{figure}

\begin{figure}[t!]
\includegraphics*[width=\linewidth]{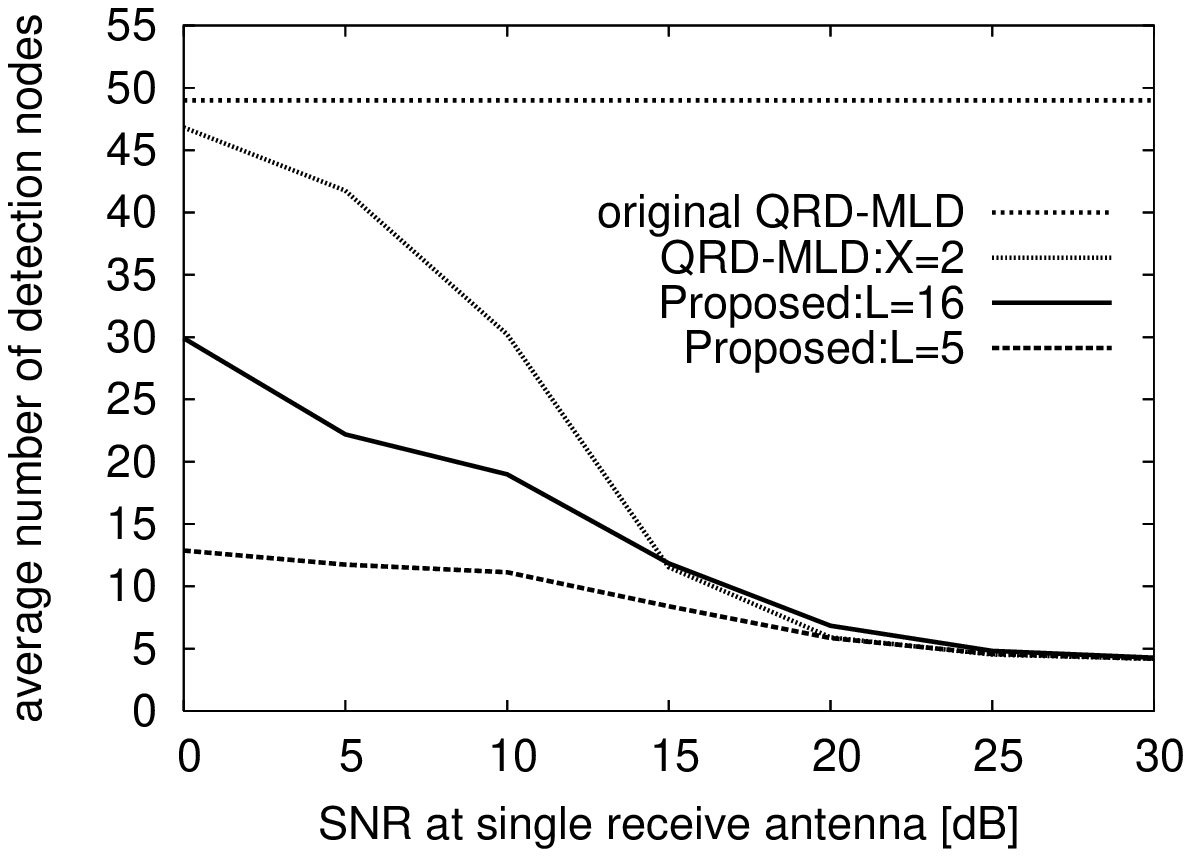}
\caption{($4 \times 4$) average number of detection nodes}
\includegraphics*[width=\linewidth]{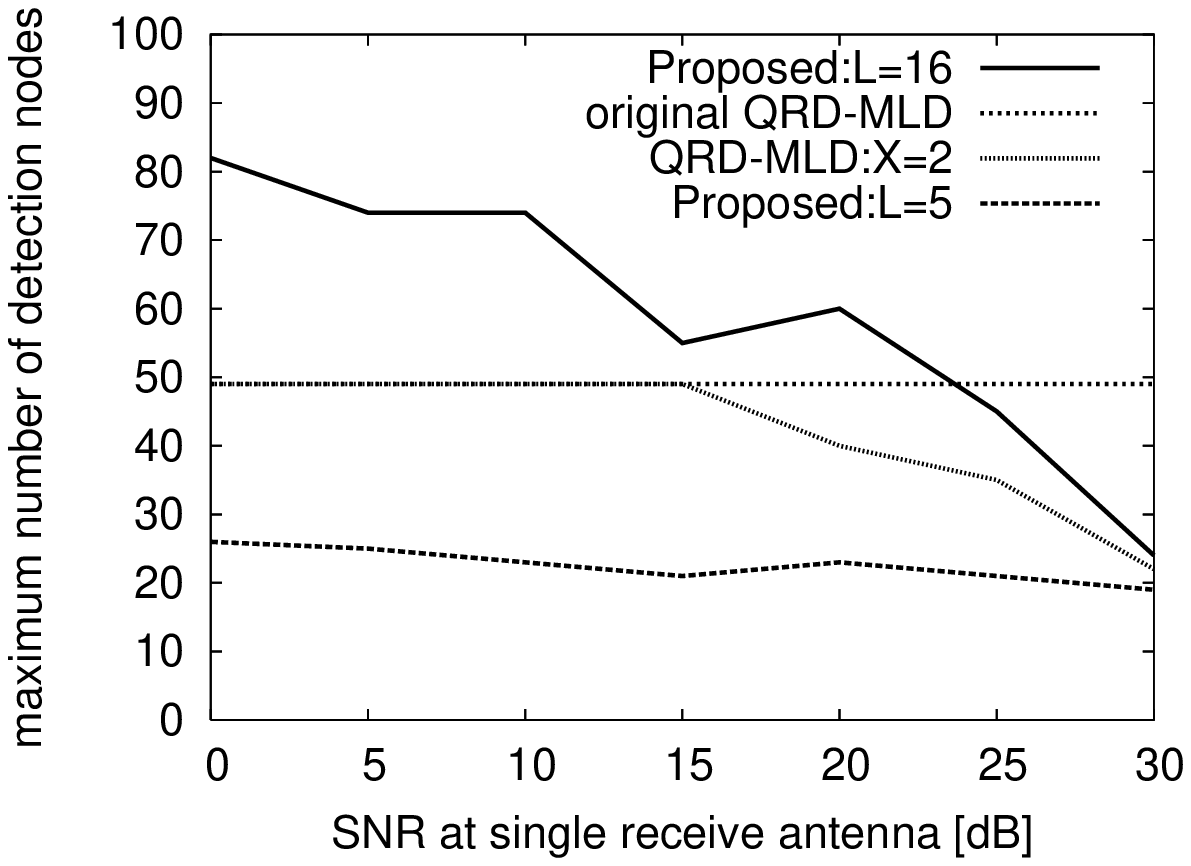}
\caption{($4 \times 4$) maximum number of detection nodes}
\end{figure}

\begin{figure}[t!]
\includegraphics*[width=\linewidth]{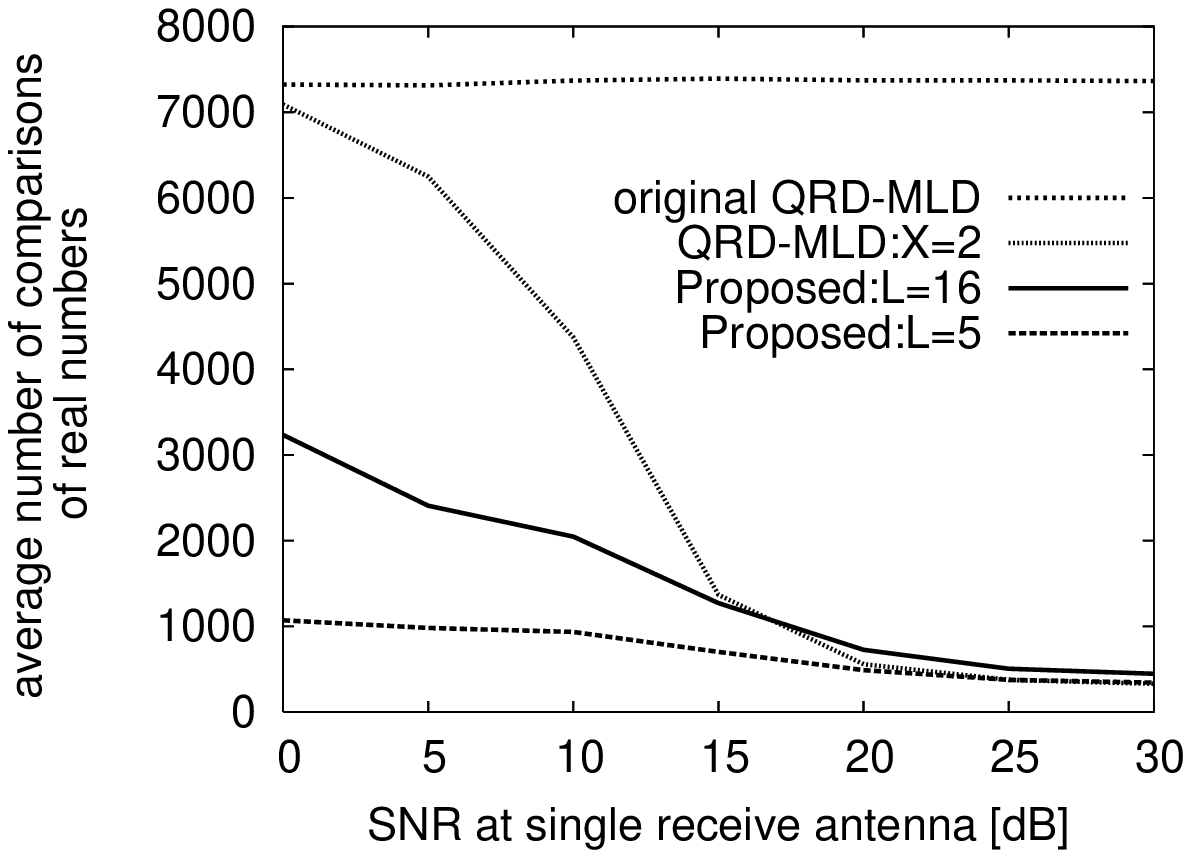}
\caption{($4 \times 4$) average number of comparisons of real numbers}
\includegraphics*[width=\linewidth]{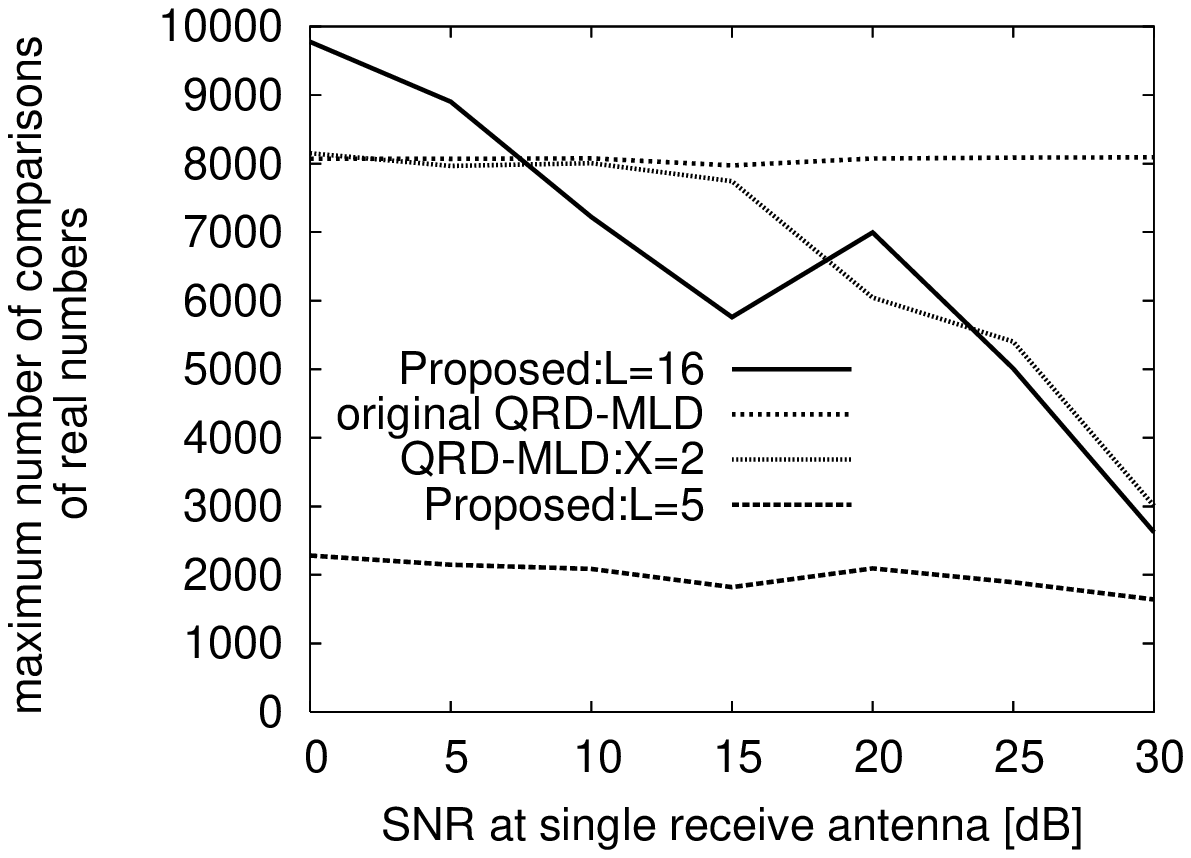}
\caption{($4 \times 4$) maximum number of comparisons of real numbers}
\end{figure}

\begin{figure}[t!]
\includegraphics*[width=\linewidth]{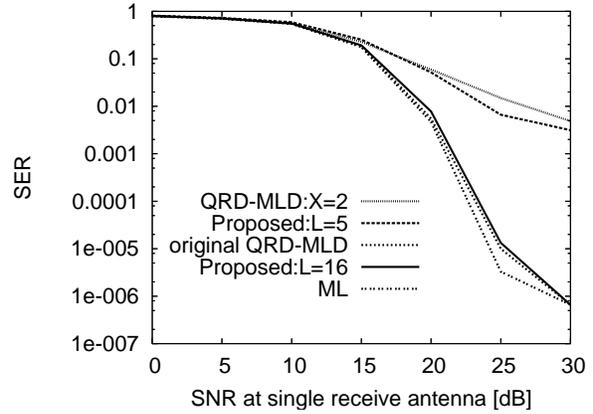}
\caption{($6 \times 6$) symbol error rate}
\end{figure}
\begin{figure}[t!]
\includegraphics*[width=\linewidth]{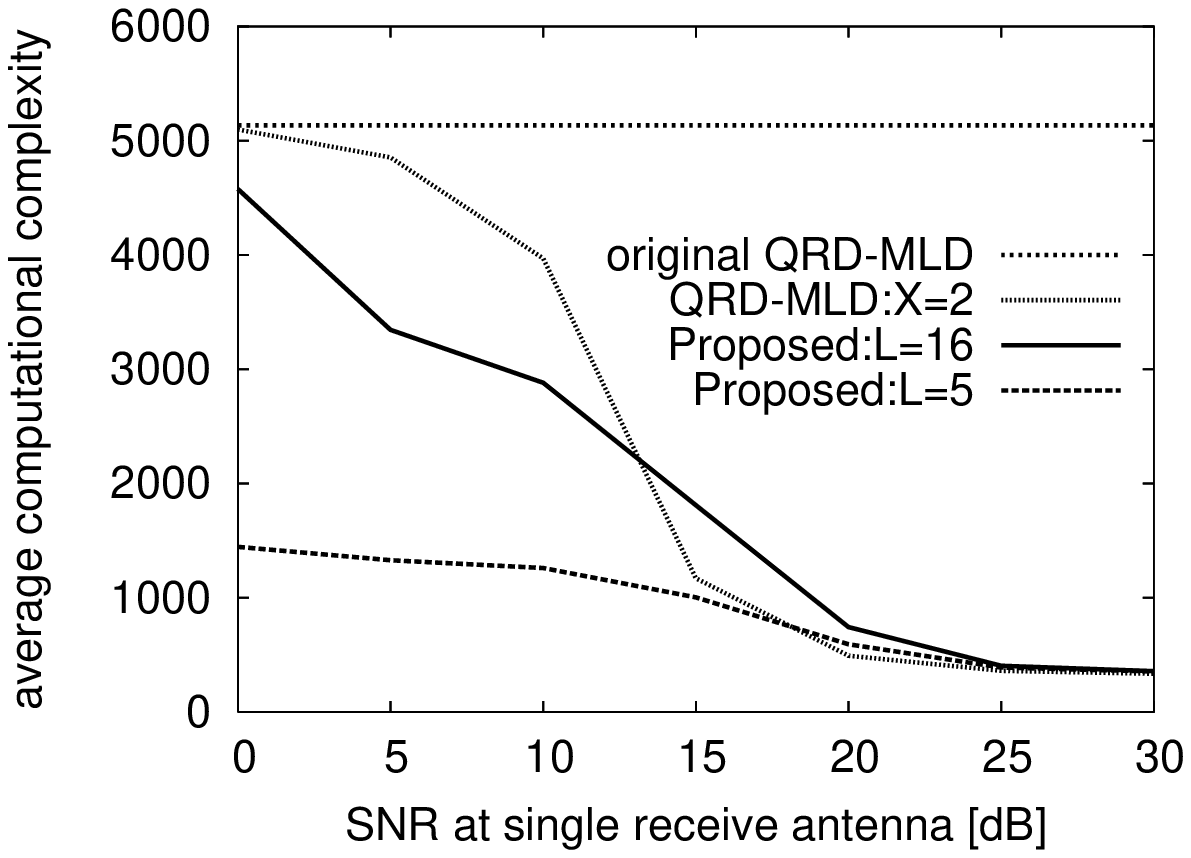}
\caption{($6 \times 6$) average computational complexity}
\includegraphics*[width=\linewidth]{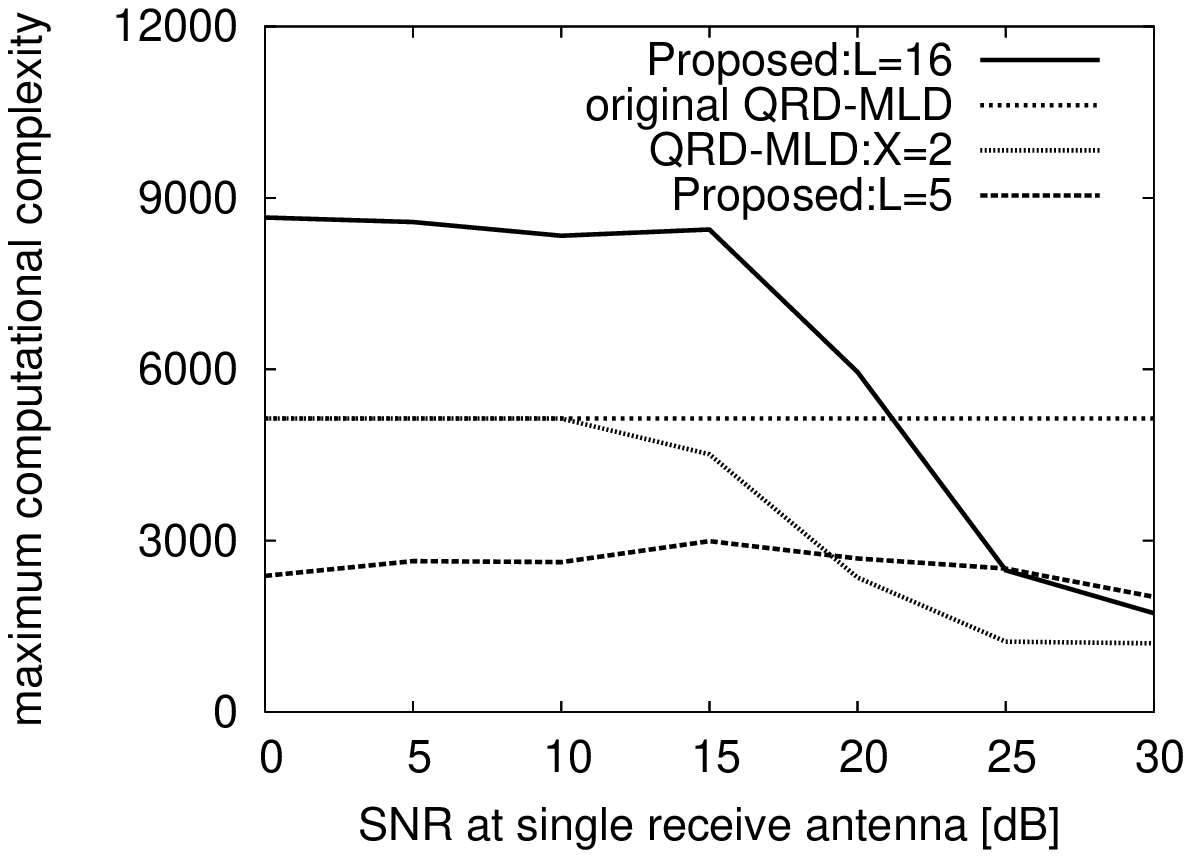}
\caption{($6 \times 6$) maximum computational complexity}
\end{figure}

\begin{figure}[t!]
\includegraphics*[width=\linewidth]{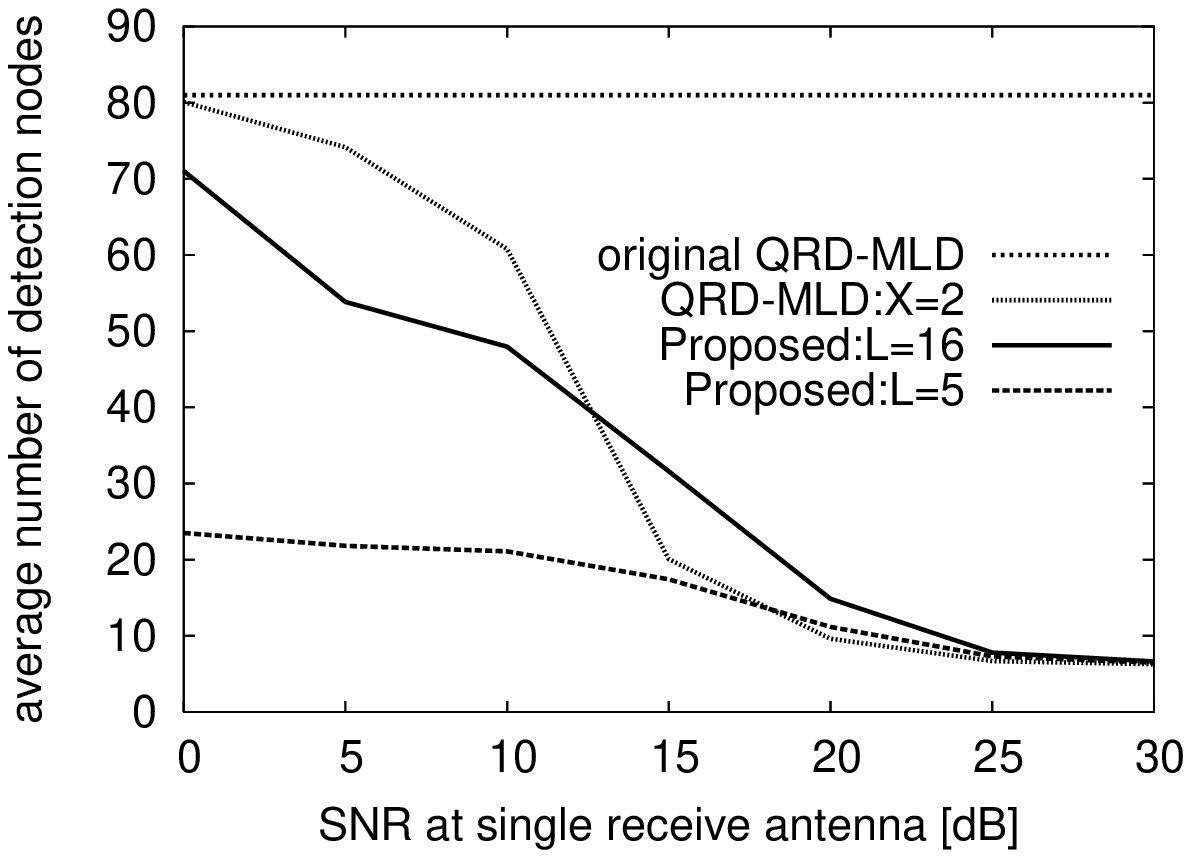}
\caption{($6 \times 6$) average number of detection nodes}
\includegraphics*[width=\linewidth]{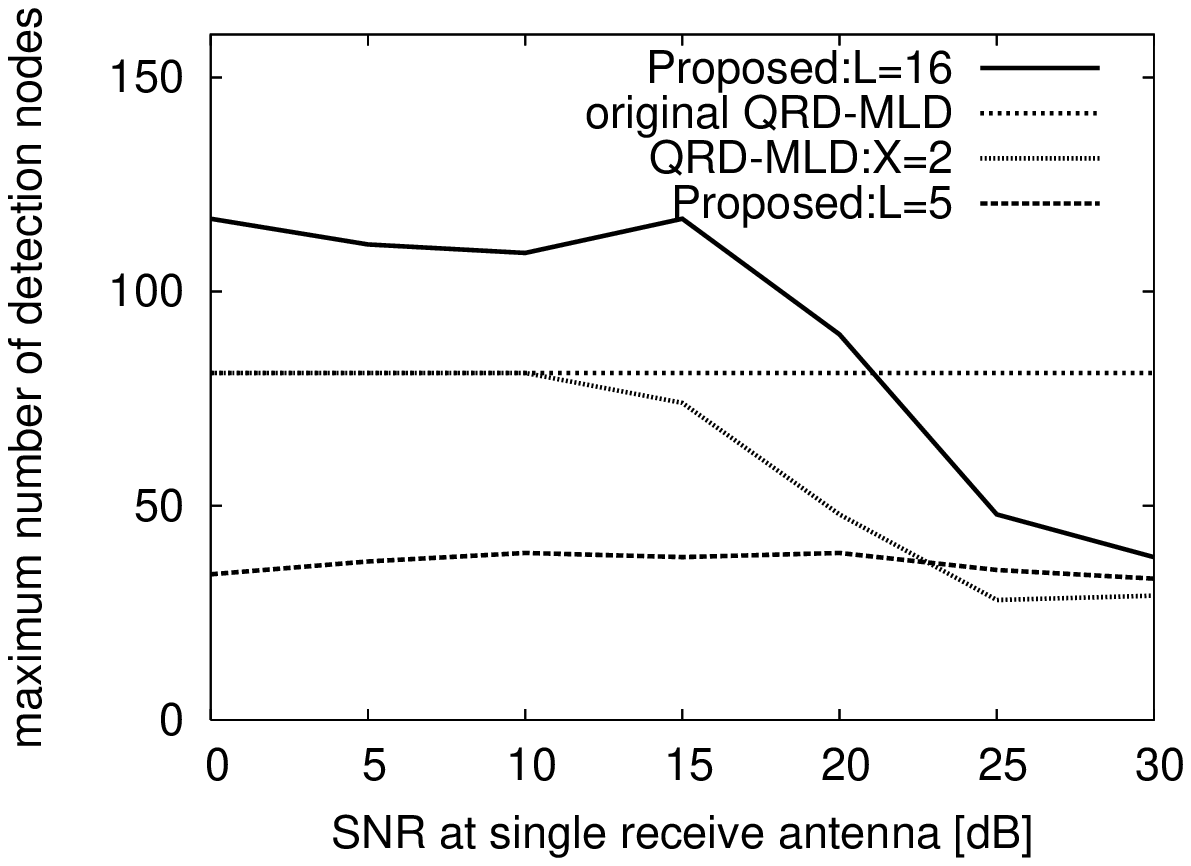}
\caption{($6 \times 6$) maximum number of detection nodes}
\end{figure}

\begin{figure}[t!]
\includegraphics*[width=\linewidth]{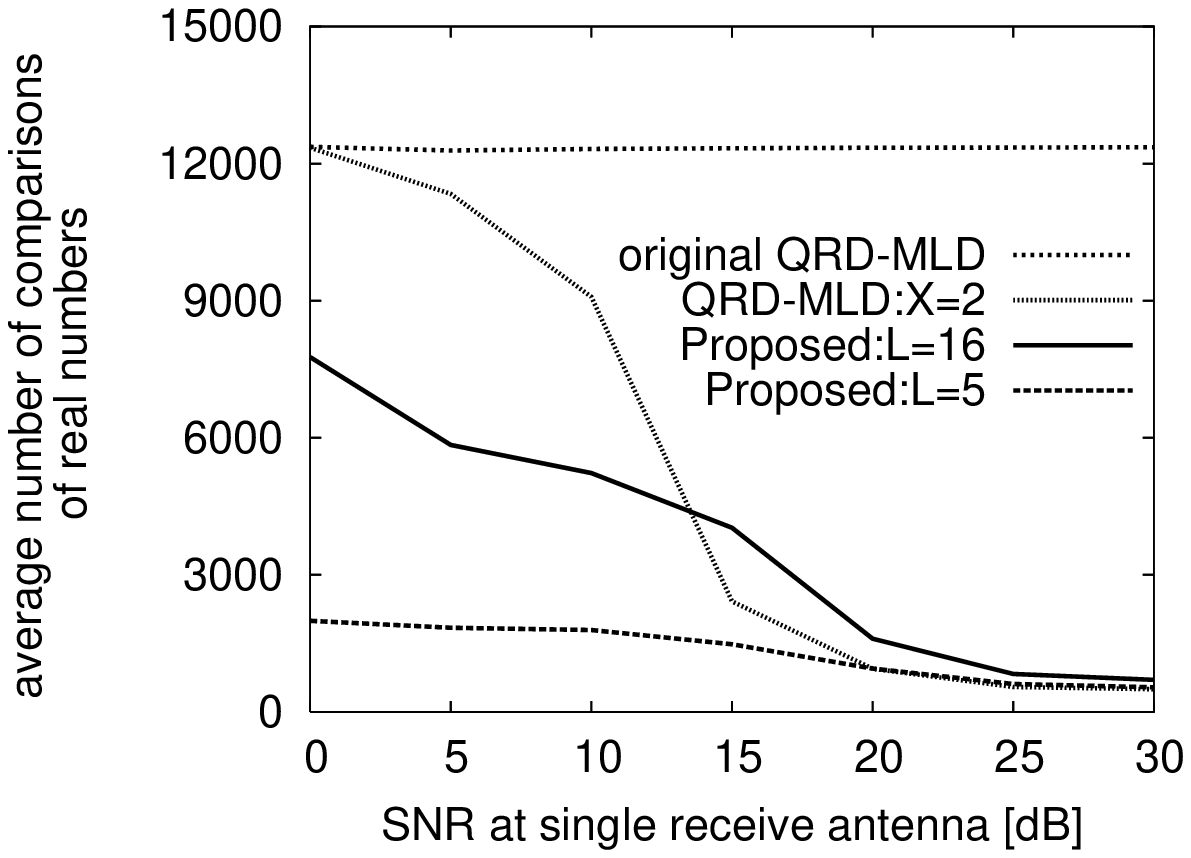}
\caption{($6 \times 6$) average number of comparisons of real numbers}
\includegraphics*[width=\linewidth]{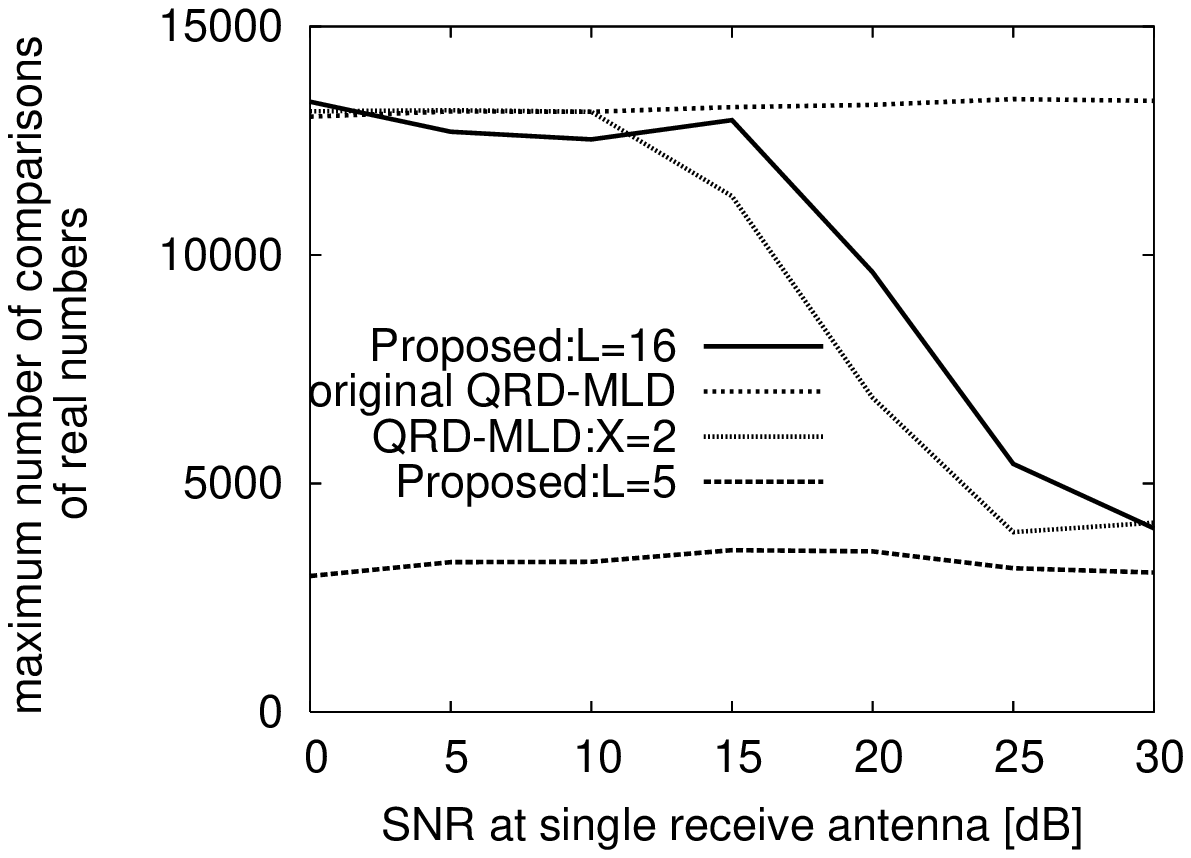}
\caption{($6 \times 6$) maximum number of comparisons of real numbers}
\end{figure}

In this section, we compare the computational complexity, 
the number of detection nodes and the number of comparisons 
of real numbers 
among the proposed algorithm and the QRD-MLDs. 
Throughout the simulations, we consider the following system model.
\begin{itemize}
\item We do two simulations. In the first simulation, the number of 
transmit antennas $t = 4$, and the number of receive antennas 
$r = 4$. In the second simulation, the number of 
transmit antennas $t = 6$, and the number of receive antennas 
$r = 6$. 
\item The signal constellation at each transmit antenna is 16-QAM 
and all signals are drawn according to the uniform i.i.d. distribution.
\item The fading coefficients obey the $CN(0,1)$ distribution, and the 
receiver knows it perfectly.
\item The noise at each recieve antenna obeys the $CN(0, \phi)$ distribution. 
$\phi$ is caluculated by $\phi^2 = t E_s \times 10^{(-SNR/10)}$, where $E_s$ is 
the average symbol energy.
\item We transmit 100000 signals, which is 400000 symbols 
if the number of transmit antennas is $4$ and 600000 symbols 
if the number of transmit antennas is $6$, and every 100 signals, 
change the fading matrix. 
\end{itemize}

If $M=16$ is used and the signal constellation is 16-QAM, QRD-MLD 
has symbol error rate (SER) near to the ML detection \cite{M}. So, we use $M=16$. 
In QRD-MLD's improvement, we use $X=2$ in (6) as used in \cite{M3}. 
In order for the proposed algorithm to have the similar SER to QRD-MLD and 
its improvement, we use two versions of proposed algorithm whose list 
sizes are $L=16$ and $L=5$. 
Figures 1 and 8 show that the proposed algorithm 
with $L=16$, the original QRD-MLD and the ML algorithm have almost the same 
SER throughout this simulations. 
The proposed algorithm with $L=5$ and QRD-MLD's 
improvement also have similar SER throughout this simulations. 

We count the number of multiplications and divisions of complex numbers as 
the computational complexity. 
Since the part of QR decomposition is the common part of all compared algorithms, 
we do not include that part in comparison of complexity.

In QRD-MLDs, 
we use the quick sort to arrange the nodes and decide $M$ nodes 
with the smallest accumulated metric value at each depth.

Because the QRD-MLD keeps $M$ nodes at each depth, the number 
of detection nodes and the computational complexity 
are completely determined by $M$. However, in the proposed algorithm and QRD-MLD's 
improvement, the number of detection 
nodes and the computational complexity are not fixed.

Figures 1--7 are the results of first simulation whose 
number of transmit antennas and receive antennas are 4. 
Figures 8--14 are the results of second simulation whose 
number of transmit antennas and receive antennas are 6. 

At first, we discuss the result of first simulation. 
According to Figures 2, 4 and 6, the propose algorithm 
with $L=16$ reduece the average computational complexity, 
average number of detection nodes and average 
number of comparisons of real numbers from the original QRD-MLD. 
Moreover, in the case of high SNR, although the proposed algorithm 
with $L=16$ has much smaller SER than QRD-MLD's improvement according to Figure 1, 
the average computational complexity 
of proposed algorithm with $L=16$ is almost the same as QRD-MLD's improvement. 
In the case of low SNR, the average computational complexity, 
average number of detection nodes and average number of 
comparisons of real numbers 
of the proposed algorithm with $L=5$ are lower than 
QRD-MLD's improvement. In the case of high SNR, the average
 computational 
complexity, average number of detection nodes and average 
number of comparisons of real numbers of 
proposed algorithm with $L=5$ are almost same 
as QRD-MLD's improvement while the proposed algorithm has smaller 
SER according to Figure 1. 
According to Figures 3, 5 and 7, in the case of low SNR, 
maximum computational complexity, 
maximum number of detection nodes and the maximum number of 
comparisons of real numbers  
of the proposed algorithm 
with $L=16$ are higher than QRD-MLDs. 
However, because the average computational complexity, 
the average number of detection nodes and 
the average number of comparisons of real number 
of the proposed algorithm 
with $L=16$ are lower than QRD-MLDs, 
we find that the proposed algorithm rarely gets high computational 
complexity, large number of detection nodes or 
large number of comparisons of real numbers. 

According to Figure 8--14, which is the result of 
second simulation, the characteristic of proposed algorithm 
dose not change with the number of antennas.

\section{Conclusion}
In this paper, we propose a near ML detection algorithm. When the list 
size is adjusted so that the
proposed algorithm has the almost same symbol error rate (SER) as the 
original QRD-MLD, 
the average of the computational complexity and the number of detection nodes 
are reduced.
When the list size is adjusted so that the
proposed algorithm has the almost same symbol error rate (SER) as the 
QRD-MLD's improvement, 
in the case of low SNR, both the average computational complexity and 
average number of detection nodes are reduced and 
in the case of high SNR, the computational 
complexity and average number of detection nodes of 
proposed algorithm is almost same 
as QRD-MLD's improvement while SER of the proposed algorithm 
becomes smaller than QRD-MLD's improvement. 

\section*{Acknowledgment}
We would like to thank Prof.\ 
Kiyomichi Araki for drawing our attention to the reference [6].
This research is partly supported by
the International Communications Foundation.

\end{document}